\documentclass[11pt]{article}
\usepackage{epsfig}
\usepackage{amsmath}
\usepackage{cite}
\def\xaa{x_{1}^{0}}
\def\xbb{x_{2}^{0}}

\def\gapprox{\lower .7ex\hbox{$\;\stackrel{\textstyle >}{\sim}\;$}}

\def\d{\hbox{d}}

\long\def\symbolfootnote[#1]#2{\begingroup%
\def\thefootnote{\fnsymbol{footnote}}\footnote[#1]{#2}\endgroup}

\begin{document}
\unitlength1cm
\begin{titlepage}
\vspace*{-1cm}
\begin{flushright}
ZU-TH 03/11
\end{flushright}
\vskip 3.5cm

\begin{center}
{\Large\bf QCD corrections to longitudinal spin asymmetries\\[2mm] in 
$W^\pm$-boson production at RHIC}
\vskip 1.cm
{\large  C.~von Arx$^{a,\, }$}\symbolfootnote[1]{present address: Swiss Federal Nuclear Safety Inspectorate, CH-5200 Brugg} and {\large T.~Gehrmann$^b$}
\vskip .7cm
{\it
$^a$ Departement Physik, Universit\"at Basel,
Klingelbergstrasse 82,
CH-4056 Basel, Switzerland
}\\[2mm]
{\it
$^b$ Institut f\"ur Theoretische Physik, Universit\"at Z\"urich,
Winterthurerstrasse 190,\\ CH-8057 Z\"urich, Switzerland}
\end{center}
\vskip 2cm

\begin{abstract}
The polarized antiquark distributions in the proton can be measured 
by studying spin asymmetries in vector boson production in 
longitudinally polarized proton-proton collisions. The STAR and PHENIX 
experiments at BNL RHIC have reported first observations of 
single spin asymmetries in $W^\pm$-production 
most recently. We compute the QCD corrections to single and double spin asymmetries, taking 
account of the leptonic decay of the $W^\pm$ boson and of restrictions on the kinematical 
acceptance of the detectors. The QCD corrections have only a small impact on the asymmetries, 
such that a reliable extraction of the polarized antiquark distributions can be envisaged once 
more precise measurements are made.
\end{abstract}
\end{titlepage}
\newpage

\section{Introduction}

Owing to a large variety of measurements, the unpolarized parton 
distributions are known to high precision~\cite{mrst,jr,abkm,cteq,nnpdf}.
Determinations of the  polarized parton distributions~\cite{dssv,others} 
have to rely 
on a much smaller data set, consisting mainly of inclusive and semi-inclusive 
polarized deep inelastic scattering measurements on fixed 
targets, augmented with 
single inclusive hadron and jet measurements from RHIC. As a result, the 
polarized quark distributions in the proton are determined accurately, 
while constraints on the polarized antiquark and gluon distributions 
are much less stringent. Semi-inclusive polarized deep inelastic scattering
provides some information on the antiquark distributions of different
flavors, although this extraction is not free from ambiguities, since it 
requires the quark-to-hadron fragmentation functions as input. 

A very clean 
measurement of the polarized antiquark distributions in the proton can be 
made from spin asymmetries in electroweak vector boson production 
in polarized proton-proton collisions~\cite{soffer}. Owing to the 
parity-violating coupling of the $W^\pm$ and $Z^0$ bosons, non-vanishing 
spin asymmetries are already obtained in the collision of one 
polarized and one unpolarized proton (single spin asymmetries), in addition to
the more common double spin asymmetries.  
Measurements of these asymmetries are possible at the BNL RHIC collider, 
and both the STAR and PHENIX experiments have recently reported 
a first observation of single spin asymmetries in $W^\pm$ 
production~\cite{star,phenix}. At present, these data are not yet sufficiently 
precise (and do not cover a large enough kinematical range) to 
provide competitive constraints on the polarized antiquark distributions. 
However, future polarized proton runs at RHIC will improve upon this situation 
and are likely to yield precision measurements of the spin asymmetries in 
$W^\pm$ production. 

To include these data in global fits of polarized parton distributions at 
next-to-leading order (NLO), the NLO QCD corrections  
must be taken into account in the evaluation of the asymmetries. 
These corrections are known already for a long time for the 
fully inclusive 
asymmetries in gauge boson production~\cite{ratcliffe,weber} and for the
asymmetries differential in the gauge boson rapidity~\cite{tgdy,tgvec}.  
Since $W^\pm$-boson production is observed only through the 
decay into a lepton and a neutrino, the $W^\pm$-boson rapidity cannot 
be measured directly. By accurately determining the 
missing transverse momentum due to the unobserved
neutrino and measuring the lepton 
transverse momentum and rapidity, the $W^\pm$-boson rapidity can 
in principle be inferred from kinematical constraints. With the limited 
coverage of the STAR and PHENIX detectors, such a determination is 
however not possible. Instead, one measures the asymmetries 
as a function of the kinematics of the observed final-state lepton: 
lepton rapidity $y_l$ and lepton transverse momentum $p_{T,l}$. These
correlate closely with the  $W^\pm$-boson rapidity, but the interpretation 
of the resulting asymmetries in terms of the polarized parton distributions is 
less straightforward. 

It is the aim of this paper to derive the NLO QCD corrections to 
the single and double spin asymmetries in $W^\pm$ production as 
function of the observed lepton rapidity and transverse momentum. For the 
single spin asymmetries, these 
corrections were recently obtained  
by de Florian and Vogelsang in~\cite{dfv}, and our calculation provides 
an independent validation of their results. Higher-order QCD effects in the 
asymmetries, including resummation of large logarithmic corrections, 
were studied previously in~\cite{rhicbos} and are implemented in the 
widely used RHICBOS program. 

This paper is structured as follows: in Section~\ref{sec:asy}, we introduce 
the single and double spin asymmetries in 
$W^\pm$-boson production and discuss their interpretation in terms of
polarized 
parton distributions. Section~\ref{sec:nlo} briefly reviews 
the technical aspects of the NLO calculation, while Section~\ref{sec:num} 
discusses the numerical impact of the NLO corrections. We compare the 
NLO results to the recent STAR and PHENIX measurements in Section~\ref{sec:rhic} and 
 conclude with 
Section~\ref{sec:conc}.

\section{Single and double spin asymmetries in vector boson production}
\label{sec:asy}
The single and double spin asymmetries in vector boson production
processes are constructed from the helicity-dependent production cross
sections
$$ \d \sigma^{h_1h_2}$$
where $h_{1,2}=(+,-)$ denote the helicities of the two incoming
hadrons. The coordinate frame is defined such that hadron 1 moves in
positive rapidity direction. 

The unpolarized cross section is the average over all initial state
helicities:
\begin{equation} \d \sigma  =  
\frac{1}{4} \left(\d \sigma^{++} + \d \sigma^{+-} 
+ \d \sigma^{-+} + \d \sigma^{--} \right)\; , \label{eq:sig0} 
\end{equation}
while the singly (only hadron 1 polarized) and doubly polarized cross sections are given by:
\begin{eqnarray}
\d \Delta \sigma_{L} & = & \frac{1}{4} \left( \d \sigma^{++} + \d \sigma^{+-} 
- \d \sigma^{-+} - \d \sigma^{--}\right)\; ,  \label{eq:sigl}\\
\d \Delta \sigma_{LL} & = & \frac{1}{4} \left( \d \sigma^{++} - \d \sigma^{+-} 
- \d \sigma^{-+} + \d \sigma^{--}\right)\; .  \label{eq:sigll}
\end{eqnarray}
From these, single and double spin asymmetries are constructed as
functions of kinematical variables of the observed final-state
particles. Previous studies on NLO corrections to 
spin asymmetries in vector boson production~\cite{tgvec}
focused on the rapidity of the vector boson $y$, which is itself not
directly measurable in $W^\pm$ production, due to the unobserved
neutrino in the final state. 

For the partonic interpretation of the asymmetries, the vector boson
rapidity is very instructive, since at leading order (LO) it is
directly related to the momentum fractions of the partons probed in
both hadrons:
$$x_{1,2}^0 = \sqrt{M_W^2/S}\; e^{\pm y}.$$
Restricting to only up and down quarks, one thus finds very simple
leading-order expressions for the asymmetries:
\begin{eqnarray}
A^{W^+}_{L} (y)& = & \frac{-\Delta u(\xaa) \bar d(\xbb) +
\Delta \bar d(\xaa) u(\xbb)}{u(\xaa) 
\bar d(\xbb) + \bar d(\xaa) u(\xbb)}\; , \nonumber \\
A^{W^-}_{L} (y)& = & \frac{- \Delta d(\xaa) \bar u(\xbb) +
\Delta \bar u(\xaa) d(\xbb)}{d(\xaa) 
\bar u(\xbb) + \bar u(\xaa) d(\xbb)}\; ,  \\
A^{W^+}_{LL} (y)& = & - \frac{\Delta u(\xaa) \Delta \bar d(\xbb) +
\Delta \bar d(\xaa) \Delta u(\xbb)}{u(\xaa) 
\bar d(\xbb) + \bar d(\xaa) u(\xbb)}\; , \nonumber \\
A^{W^-}_{LL} (y)& = & - \frac{\Delta d(\xaa) \Delta \bar u(\xbb) +
\Delta \bar u(\xaa) \Delta d(\xbb)}{d(\xaa) 
\bar u(\xbb) + \bar u(\xaa) d(\xbb)}\; . 
\end{eqnarray}
At RHIC with $\sqrt{S}=500$~GeV, these asymmetries are thus sensitive on 
$x_{1,2} \gapprox 0.05$.
At large and positive rapidity $y$, ($\xaa > \xbb$), the single 
spin asymmetries $A_L$ are dominated by the first term in the numerator, since 
$\Delta q(x_1^0,Q^2) \gg \Delta \bar q(x_1^0,Q^2)$ for large 
$x_1^0$. The second term in the numerator
is dominant for large and negative $y$, ($\xaa < \xbb$), since 
$q (x_2^0,Q^2) \gg \bar q(x_2^0,Q^2)$ for large $x_2^0$.  
Given that for the $x$-range probed by RHIC, the polarized quark
distributions are substantially larger than the polarized antiquark
distributions, and their sum is well constrained from polarized
inclusive deep inelastic scattering~\cite{dssv}, the double spin
asymmetries can be used for a reliable  extraction of $\Delta \bar u(x)$
and $\Delta \bar d(x)$. 

With only the decay lepton being
observable, the lepton rapidity $y_l$ and transverse momentum $p_{T,l}$
are the more appropriate kinematical variables. Measurements of the
transverse mass, as commonly carried out in vector boson production at
the Tevatron, are not feasible at RHIC due to the non-hermetic
detector coverage, which prevents a reliable reconstruction of the
missing transverse momentum carried by the neutrino. 

We shall thus focus on the single and double spin asymmetries
\begin{equation}
A_L(y_l) \equiv  \frac{{\d \Delta \sigma_{L} / \d y_l}} {
{\d \sigma}/{\d y_l}},\qquad 
A_{LL}(y_l) \equiv  \frac{{\d \Delta \sigma_{LL}}/{\d y_l}}{
{\d \sigma}/{\d y_l}}
\end{equation}
and
\begin{equation}
A_L(p_{T,l}) \equiv  \frac{{\d \Delta \sigma_{L}}/{\d p_{T,l}}}{
{\d \sigma}/{\d y_{T,l}}},\qquad 
A_{LL}(p_{T,l}) \equiv  \frac{{\d \Delta \sigma_{LL}}/{\d p_{T,l}}}{
{\d \sigma}/{\d p_{T,l}}}\;.
\end{equation}
Polarized and unpolarized cross sections in these asymmetries can 
be expanded in QCD perturbation theory in the strong 
coupling constant $\alpha_s$. In these cross sections, the 
exchanged $W^\pm$ boson is not required to be on-shell, its finite 
width is taken into account in its propagator.

\section{Next-to-leading order corrections}
\label{sec:nlo}
The next-to-leading order expressions for the asymmetries 
are obtained by including the ${\cal O}(\alpha_s)$ corrections in numerator and denominator. 
The computation of these corrections is a standard task in higher order calculations, 
carried out in dimensional regularization in $d=4-2\epsilon$ dimensions. It requires 
to combine the virtual one-loop corrections to the process $q\bar q'\to W^{\pm}\to l^{\pm}\nu$ with the 
real radiation contributions $q\bar q' \to  W^{\pm}g\to l^{\pm}\nu g$ and $qg\to W^{\pm}q'\to l^{\pm}\nu q'$.
The real radiation contributions develop infrared singularities if the final-state gluon becomes 
soft, or if the final-state parton becomes collinear with one of the incoming partons. These 
infrared real radiation singularities can be extracted analytically. They cancel in the 
final expression for the cross section when
combined with the 
infrared-divergent virtual contributions and with mass factorization counterterms for the 
incoming parton distributions. Various techniques for the treatment of infrared singular 
real radiation exist at NLO, based either on introducing subtraction terms~\cite{cs,fks,ant} 
or on a slicing of the final-state phase space~\cite{gg,ggk}. 

In our calculation, we 
use the phase space slicing method~\cite{gg,ggk} for the real radiation contributions. In this method,
the real radiation phase space is split into resolved and unresolved regions by 
 introducing a cut-off parameter $s_{{\rm min}}$ on all 
Mandelstam invariants. The resolved regions are integrated numerically, while the unresolved regions 
are integrated analytically, giving rise to the infrared divergent terms. The phase space slicing 
contributions from the unresolved regions are universal and depend only on the type of partons 
involved in the unresolved configuration~\cite{gg}. These are computed first in 
a hypothetical kinematical situation with all partons in the final state, and then 
continued to the true kinematical situation by using a crossing function~\cite{ggk} for 
each initial state parton. The polarized crossing functions are obtained from the 
unpolarized expressions in~\cite{ggk} by substituting the $d$-dimensional
unpolarized one-loop splitting functions by their polarized counterparts~\cite{mvn,wv}.

Our calculation was validated by comparing the NLO unpolarized
differential cross sections $\d \sigma$ against MCFM~\cite{mcfm} and
the singly polarized differential cross sections $\d \Delta \sigma_L$
against CHE~\cite{dfv} (using the same set of parameters
 and setting renormalization and factorization scales 
$\mu^2 = M_W^2$ for the total cross 
sections and $\mu^2 = (M_W^2+p_{T,l}^2)/4$ for 
the distributions, as in~\cite{dfv}). 
Full agreement is found for both quantities,
thereby providing a crucial independent cross-check of~\cite{dfv}. 
\begin{figure}[t]
\begin{center}
\epsfig{file=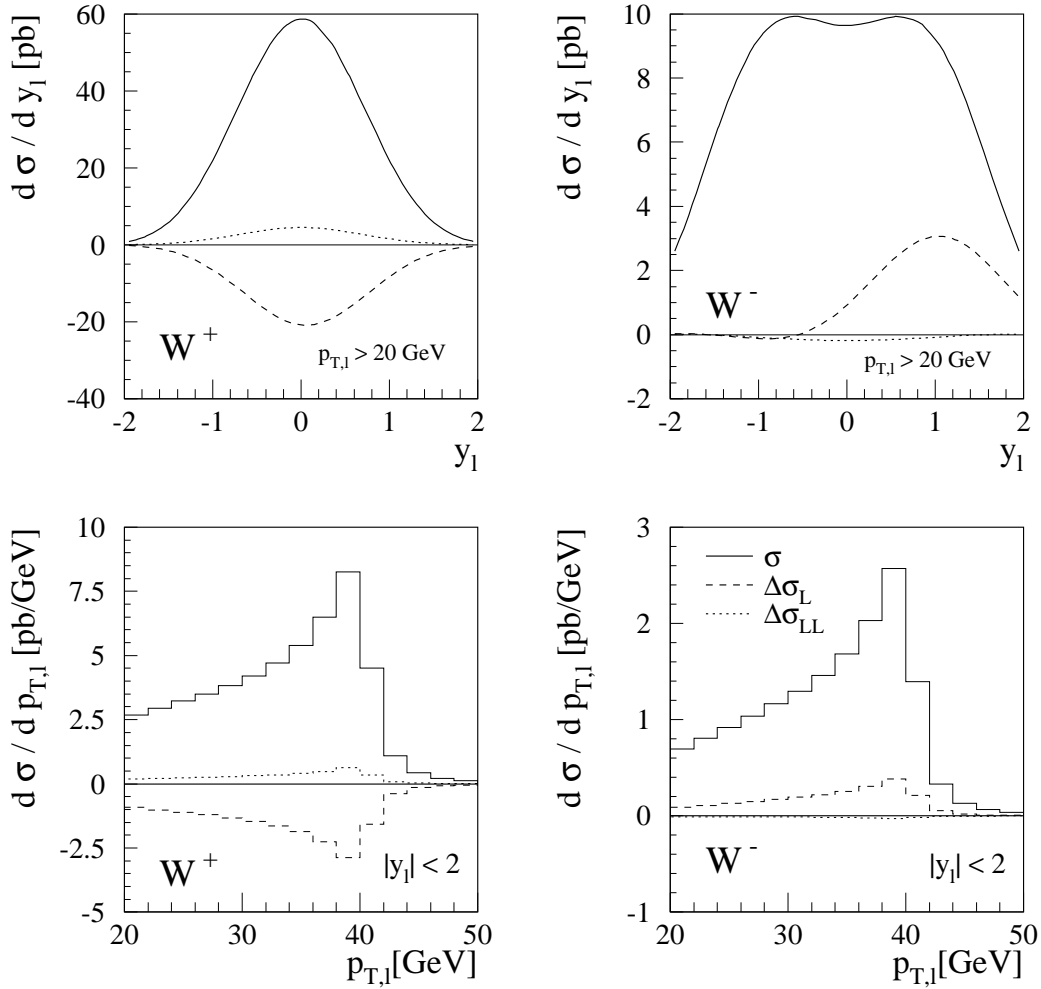,angle=-90,width=14cm}
\end{center}
\caption{Next-to-leading order 
differential cross sections for single inclusive lepton 
production at RHIC, $\sqrt{S} = 500$~GeV with $p_{T,l}>20$~GeV and $|y_l|<2$. 
Solid: unpolarized cross section, dashed: singly polarized, dotted: 
doubly polarized.\label{fig:sig}}
\end{figure}

\section{Numerical results}
\label{sec:num}
Polarized proton-proton collisions are studied at RHIC at
different centre-of-mass energies. For vector boson production, only
the high-energy runs at $\sqrt{S} = 500$~GeV are of relevance, such
that we use this setup for our studies. For the unpolarized parton
distribution functions, we use the MRST2002 NLO set~\cite{mrst02};
polarized parton distributions are taken from DSSV~\cite{dssv}. The
DSSV set is the result of a global fit to spin
asymmetries in inclusive and semi-inclusive deep inelastic
scattering as well as single inclusive hadron and jet production at
RHIC. 
In its extraction, the unpolarized cross sections entering
the asymmetries are computed from the MRST2002 NLO set, such that the
DSSV spin-dependent parton distributions should be used consistently
with this set. Likewise, the value of the strong coupling constant
should be consistent with what is used in the parton distributions,
such that we take $\alpha_s(M_Z) = 0.1195$~\cite{mrst02}.
Renormalization and factorization scales are taken to be equal, and
are fixed to $\mu_R=\mu_F=M_W$. The NLO scale dependence of the single and
double spin asymmetries $A_{L}(y)$ and $A_{LL}(y)$ was studied in
detail in~\cite{tgvec}. It turns out to be very small,
largely due to cancelations between the polarized and unpolarized
cross sections in the asymmetry.

The $W^\pm$ mass and width are taken~\cite{pdg} as $M_W = 80.399$~GeV and
$\Gamma_W=2.085$~GeV, and the Fermi coupling as 
$G_F=1.16637\cdot 10^{-5}$~GeV$^{-2}$. By using the Fermi coupling
constant as input parameter, electroweak corrections to the cross
sections are minimized.
 Neglecting incoming bottom quarks, we use the
CKM matrix elements $|V_{ud}|=|V_{cs}|=0.975$ and
$|V_{us}|=|V_{cd}|=0.222$. 

As default for our numerical studies, we use the following 
cuts on the final-state lepton: $p_{T,l} > 20$~GeV and
 $|\eta_l|<2$. The unpolarized and polarized 
NLO cross sections with these cuts are
 displayed in Figure~\ref{fig:sig} as function of either $y_l$ or 
$p_{T,l}$. From these results, it can be anticipated that the 
single spin asymmetries are substantially larger
than the double spin asymmetries. The transverse momentum 
distribution of the lepton shows the characteristic peak at 
$M_W/2$ in the unpolarized and polarized cases. To study the 
behavior of the polarized cross sections in more detail, the 
single and double spin asymmetries $A_L$ and $A_{LL}$ are more suitable 
than the cross sections themselves. 
\begin{figure}[t]
\begin{center}
\epsfig{file=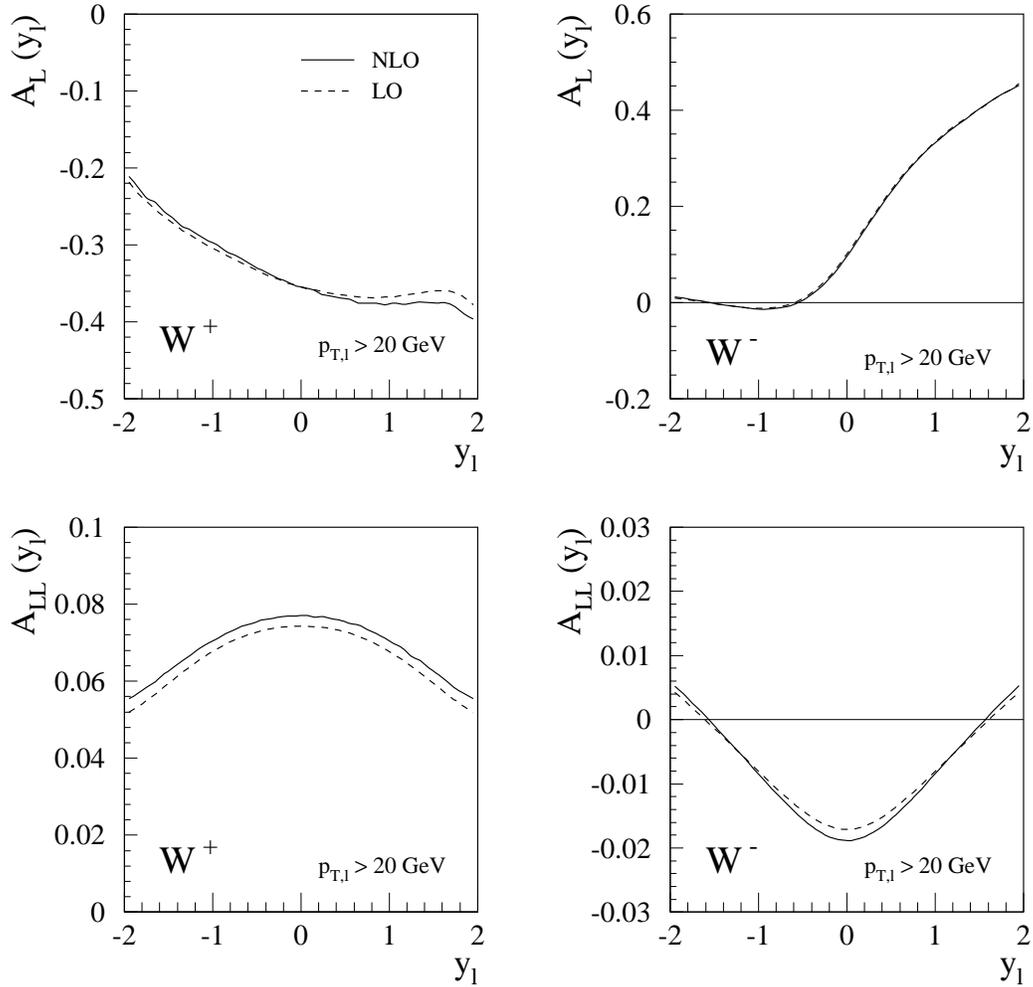,angle=-90,width=14cm}
\end{center}
\caption{Single and double spin asymmetries as function of $y_l$ at 
NLO (solid) and LO (dashed).
\label{fig:eta_all}}
\end{figure}
\begin{figure}[t]
\begin{center}
\epsfig{file=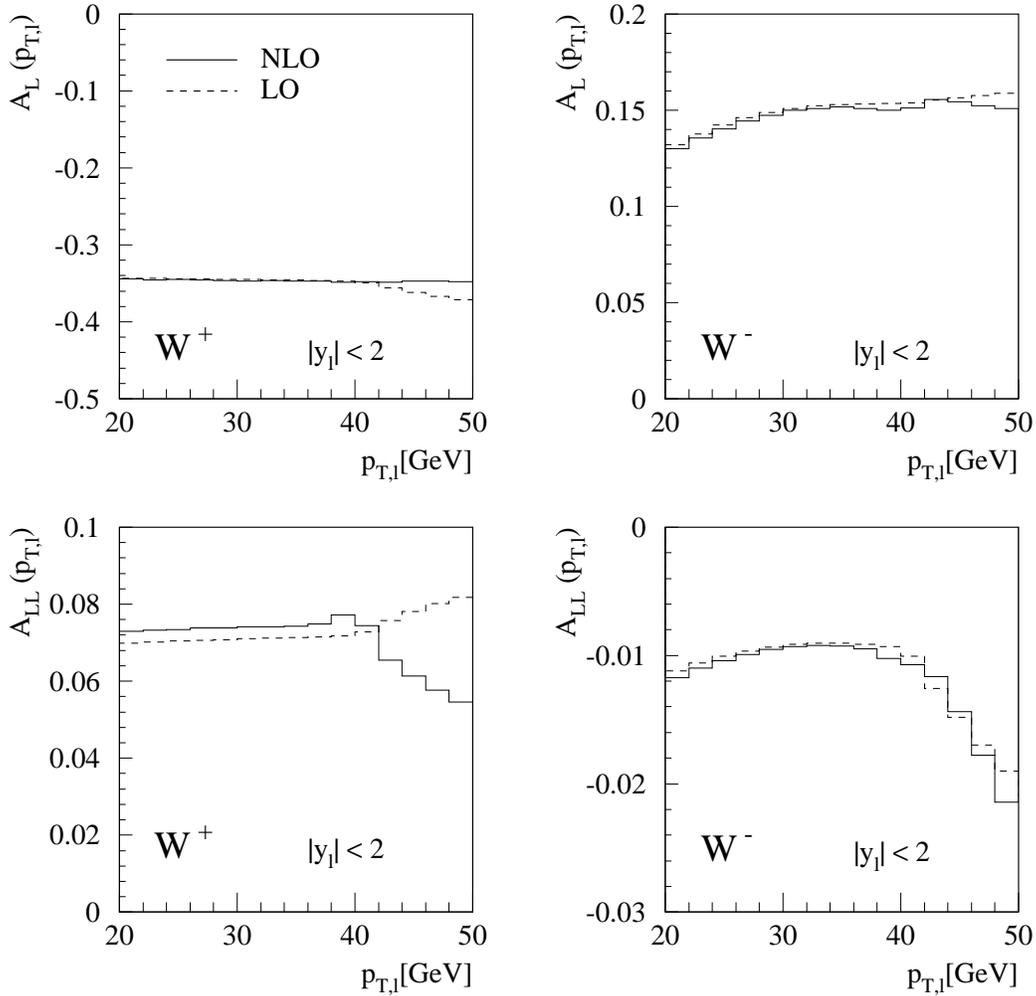,angle=-90,width=14cm}
\end{center}
\caption{Single and double spin asymmetries as function of $p_{T,l}$ at 
NLO (solid) and LO (dashed).
\label{fig:pt_all}}
\end{figure}

Figures~\ref{fig:eta_all} and \ref{fig:pt_all} show these asymmetries 
as function of $y_l$ and $p_{T,l}$. In these figures, the LO asymmetry is 
obtained by evaluating polarized and unpolarized cross sections to LO, while 
both cross sections are evaluated to NLO for the NLO asymmetry. Since 
this figure aims to quantify the effect of the parton-level NLO corrections, 
both LO and NLO asymmetry are evaluated with the same NLO sets of 
parton distributions. We observe that  $A_L(y_l)$ and $A_{LL}(y_l)$ are 
only very mildly affected by the NLO corrections, and that the 
corrections do not change the shape of these asymmetries.  The corrections 
to  $A_L(p_{T,l})$ and $A_{LL}(p_{T,l})$ are also very small and 
uniform for $p_{T,l}<M_W/2$, while they become more pronounced for 
 $p_{T,l}>M_W/2$. This behavior can be easily understood from kinematical 
considerations. At LO, the $W^\pm$ boson is produced at vanishing 
transverse momentum, such that leptons above  $p_{T,l}=M_W/2$ must come from 
off-shell $W^\pm$ production. The real radiation contributions at NLO 
lead to final states with a finite transverse momentum of the $W^\pm$, which 
can thus be near its mass shell. With this new 
parton-level contribution appearing at NLO, the NLO 
unpolarized and polarized cross sections for  $p_{T,l}>M_W/2$ are much larger 
than their LO counterparts, and the asymmetry is modified substantially.  

Compared to the results of~\cite{dfv}, we observe slight discrepancies 
in $A_L(y_l)$ for large $|y_l|$. These are entirely due to the different 
choice of renormalization and factorization scales: $\mu^2=M_W^2$ in 
our evaluation, compared to $\mu^2=(M_W^2+p_{T,l}^2)/4$ in~\cite{dfv}. With the 
same scale choice (and the same 
electroweak parameters), we find full agreement with~\cite{dfv}. 
\begin{figure}[t]
\begin{center}
\epsfig{file=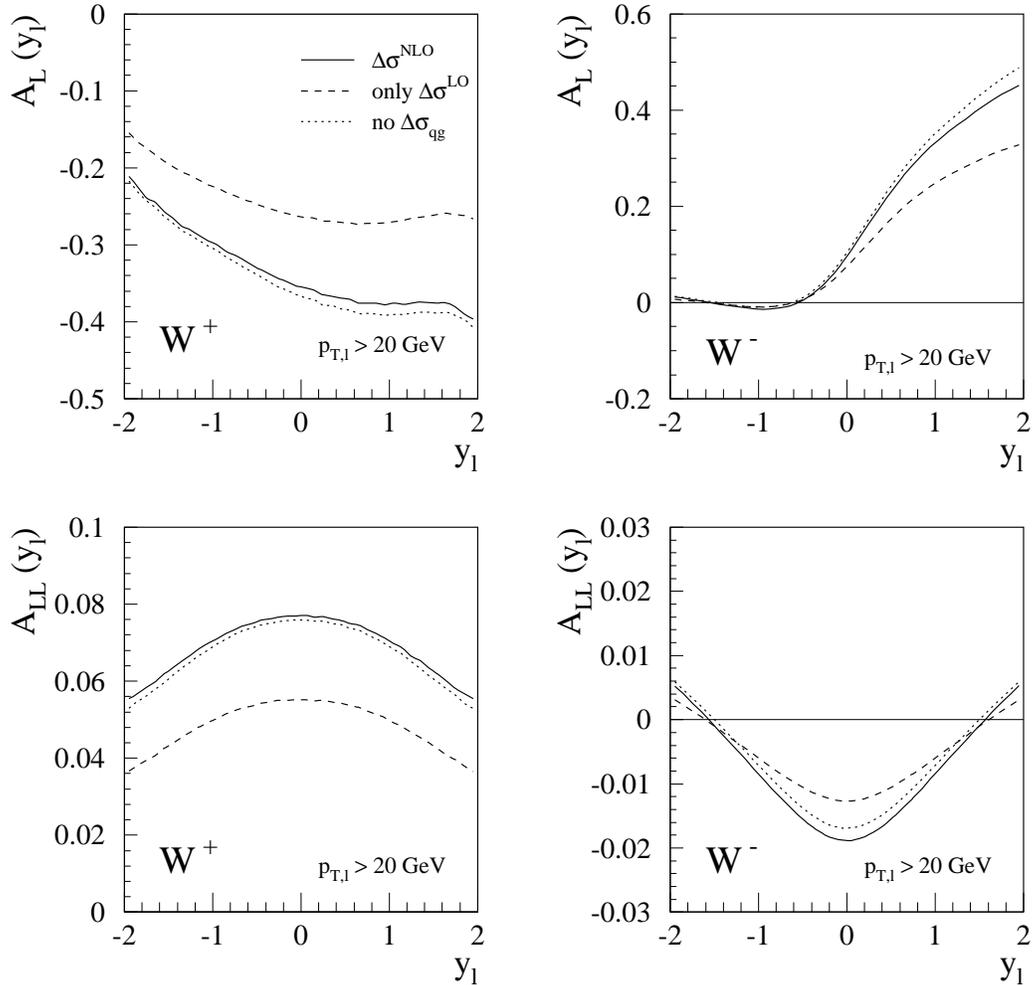,angle=-90,width=14cm}
\end{center}
\caption{Individual partonic contributions to the single and double 
spin asymmetries as function of $y_l$. Unpolarized cross section is always 
evaluated to NLO, 
polarized cross section evaluated to NLO (solid), LO (dashed), NLO without
$qg$ subprocess (dotted).
\label{fig:eta_c}}
\end{figure}
\begin{figure}[t]
\begin{center}
\epsfig{file=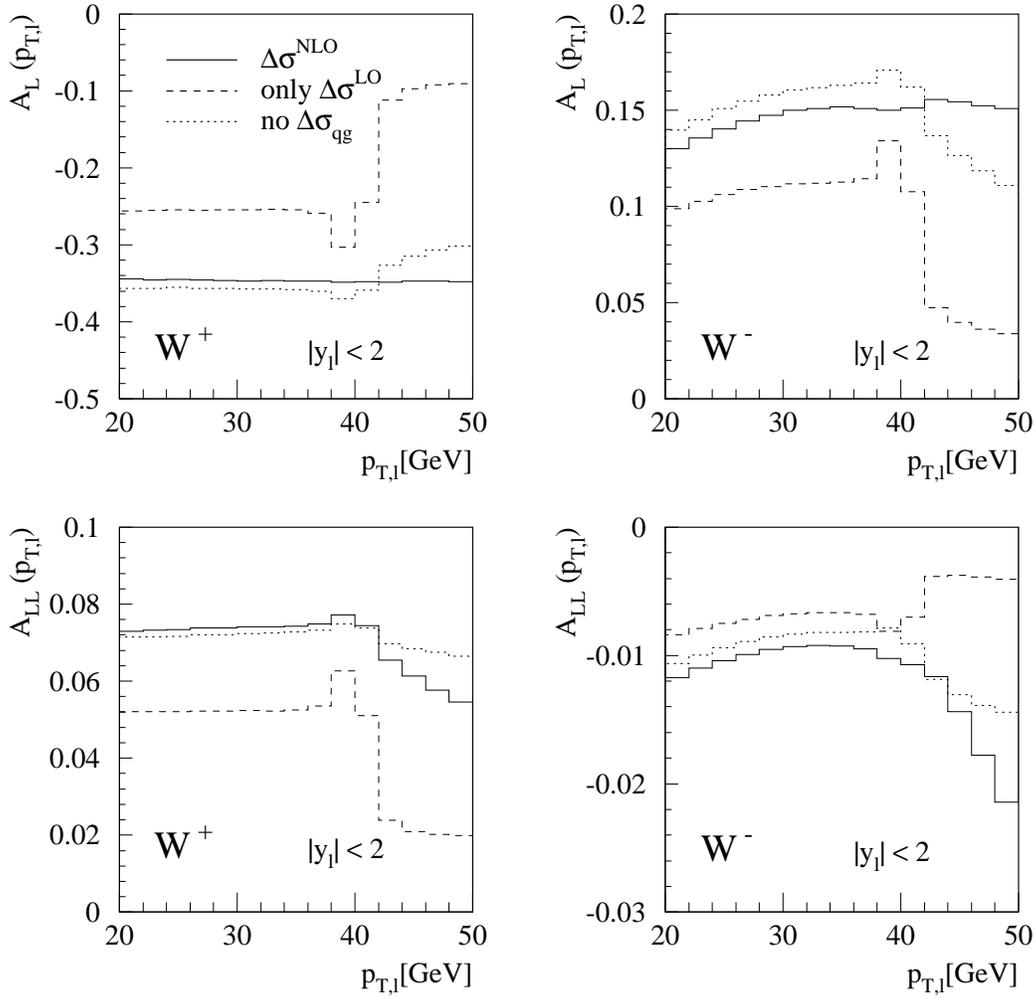,angle=-90,width=14cm}
\end{center}
\caption{Individual partonic contributions to the single and double 
spin asymmetries as function of $p_{T,l}$. Unpolarized cross section is always 
evaluated to NLO, 
polarized cross section evaluated to NLO (solid), LO (dashed), NLO without
$qg$ subprocess (dotted).
\label{fig:pt_c}}
\end{figure}
 
The numerical importance of the different parton-level 
contributions to the single and double spin asymmetries is displayed in 
Figures~\ref{fig:eta_all} and~\ref{fig:pt_all}. All asymmetries in these 
figures are normalized to the unpolarized cross sections evaluated at 
NLO. Comparing the results obtained with $\Delta \sigma$ evaluated only at 
LO with the LO asymmetries in Figures~\ref{fig:eta_all} and~\ref{fig:pt_all}, 
one concludes that the smallness of the NLO corrections to the asymmetries 
is due to a cancelation of the NLO corrections to unpolarized and polarized 
cross sections in the asymmetry. The NLO corrections to the polarized cross sections are
moreover dominated by the $q\bar q$ subprocess. Consequently, a future extraction 
of polarized quark distributions from the $A_L(y_l)$ and $A_{LL}(y_l)$ will not be 
affected by the uncertainty on the polarized gluon distribution.

\section{Comparison with RHIC data}
\label{sec:rhic}
The single spin asymmetries $A_L$ in $W^\pm$-boson production have been measured recently 
at RHIC by the STAR~\cite{star} and PHENIX~\cite{phenix} experiments. 
Both measurements are based on detecting only electrons and positrons at large 
transverse momentum. Cuts are imposed on the lepton transverse momentum and 
rapidity, and the measurement is performed in a single bin. Especially the 
PHENIX measurement relies on a very limited kinematical coverage, such that 
$W^\pm$ and $Z^0$ boson production cannot be disentangled, since the second 
lepton from the $Z^0$ decay is often outside the coverage. The $Z^0$-boson 
contributions to the single spin asymmetries at NLO were investigated 
in~\cite{dfv}. 
\begin{table}[t]
\begin{tabular}{|l|l|l|r|r|r|}\hline
\rule[-4mm]{0mm}{10mm}& Cuts & & $A_L$ (exp) & $A_L$ (LO) & $A_L$ (NLO)\\[2mm] \hline \hline
\rule[-4mm]{0mm}{10mm} STAR & 25 GeV $< p_{T,l} <$ 50 GeV; & $W^+$ & $-0.27 \pm 0.10 \pm 0.02$ & -0.347 & -0.348\\[2mm] \cline{3-6}
\rule[-4mm]{0mm}{10mm} &  $|y_l| < 1$ & $W^-$ & $0.14\pm 0.19\pm 0.02$ & 0.127 & 0.123 \\[2mm] \hline
\rule[-4mm]{0mm}{10mm} PHENIX & 30 GeV $< p_{T,l} <$ 50 GeV; & $W^+$ & $-0.86^{+0.30}_{-0.14}$ & -0.342 & -0.342\\[2mm] \cline{3-6}
\rule[-4mm]{0mm}{10mm}& $|y_l| < 0.35$ & $W^-$ & $0.88^{+0.12}_{-0.71}$ & 0.107 & 0.103 \\[2mm] \hline
\end{tabular}
\caption{Single spin asymmetries measured in $W^\pm$ production at RHIC by STAR~\protect{\cite{star}} and 
PHENIX~\protect{\cite{phenix}} compared to LO and NLO QCD predictions.\label{tab:one}}
\end{table}

In Table~\ref{tab:one}, we summarize the experimental cuts and compare 
the experimentally measured 
asymmetries with the theoretical predictions at LO and NLO. We observe 
that the predictions for the asymmetries
are perturbatively very stable, with practically 
no shift observed for the $W^+$ production, and only a modification of 
four per cent for the $W^-$ production. The asymmetries observed 
by STAR are in good agreement with the theoretical prediction, 
while PHENIX typically obtains asymmetries with a larger magnitude than 
expected, although with large errors. Both measurements 
have clearly demonstrated the existence of single spin asymmetries 
in $W^\pm$ production, but 
not yet attained  sufficient precision to provide 
meaningful constraints on the polarized antiquark distributions.

Future more precise data on the single spin asymmetries, as well as first 
data on the double spin asymmetries will provide very important input 
to the determination of the polarized parton distributions.

\section{Conclusions}
\label{sec:conc}
Spin asymmetries in gauge boson production in polarized proton-proton 
collisions are currently measured at 
RHIC~\cite{star,phenix}. These asymmetries provide a direct
measurement of the polarized light antiquark distributions, which were
only loosely constrained from semi-inclusive polarized deep inelastic
scattering up to now~\cite{dssv,others}.
In this paper, we derived the NLO QCD corrections to single and double
spin asymmetries in $W^\pm$ boson production in polarized
proton-proton collisions as function of the measurable final-state
lepton variables. Our results for the single spin asymmetries confirm
a recent calculation of de Florian and Vogelsang~\cite{dfv}. 

We show
that the NLO corrections have only a very moderate 
impact on the asymmetries, and that the corrections are largely 
independent on the at present poorly constrained polarized gluon distribution. 
Their inclusion is nevertheless mandatory in 
 future global NLO
fits of polarized parton distributions for the sake of internal 
perturbative consistency.

\section*{Acknowledgements}

We would like to thank Daniel de Florian and 
Werner Vogelsang for help in comparing with the results
presented in~\cite{dfv}.
This research was supported by the Swiss National Science Foundation
(SNF) under contracts 200020-126691 and 200020-122287.

\end{document}